\documentclass[pdftex,twocolumn,epjc3]{svjour3}          
\RequirePackage[T1]{fontenc}
\smartqed

\RequirePackage{graphicx}
\RequirePackage{mathptmx}      
\RequirePackage{flushend}
\usepackage{amsmath,latexsym}
\RequirePackage[numbers,sort&compress]{natbib}
\RequirePackage[colorlinks,citecolor=blue,urlcolor=blue,linkcolor=blue]{hyperref}
\DeclareMathOperator{\Tr}{Tr}
\journalname{Eur. Phys. J. C}
\begin{document}

\title{General approach to the Lagrangian ambiguity in $f(R,T)$ gravity} 


\author{G. A. Carvalho\thanksref{addr1, e1}
  \and F. Rocha \thanksref{addr2, e2}
  \and H. O. Oliveira \thanksref{addr2}
  \and R. V. Lobato\thanksref{addr3, e3}
}

\thankstext{e1}{e-mail: geanderson.araujo.carvalho@gmail.com}
\thankstext{e2}{e-mail: rocha.pereira.flavia@gmail.com}
\thankstext{e3}{e-mail: ronaldo.lobato@tamuc.edu}
\institute{Instituto de Pesquisa e Desenvolvimento (IP\&D), Universidade do Vale do Para\'\i ba, 12244--000, S\~ao Jos\'e dos Campos, SP, Brazil\label{addr1}
\and Departamento de F\'\i sica, Instituto Tecnol\'ogico de Aeron\'autica, S\~ao Jos\'e dos Campos, SP, 12228--900, Brazil\label{addr2}
  \and
          Department of Physics and Astromomy, Texas A\&M University-Commerce, Commerce, TX 75428, USA\label{addr3}}

\date{Received: date / Accepted: date}

\maketitle
\begin{abstract}
The $f(R,T)$ gravity is a theory whose gravitational action depends arbitrarily on the Ricci scalar,
$R$, and the trace of the stress-energy tensor, $T$; its field equations also
depend on matter Lagrangian, $\mathcal{L}_{m}$. In the modified theories of gravity where field equations
depend on Lagrangian, there is no uniqueness on the Lagrangian definition and the dynamics of
the gravitational and matter fields can be different depending on the choice performed. In this work,
we have eliminated the $\mathcal{L}_{m}$ dependence from $f(R,T)$ gravity field equations by
generalizing the approach of Moraes in Ref.~\cite{Moraes2019}. We also propose a general approach where we
argue that the trace of the energy-momentum tensor must be considered an ``unknown'' variable of the field
equations. The trace can only depend on fundamental constants and few inputs from the standard
model. Our proposal resolves two limitations: first the energy-momentum tensor of the
$f(R,T)$ gravity is not the perfect fluid one; second, the Lagrangian is not well-defined. As a test of our approach we applied it to the study of the matter era in cosmology, and the theory can successfully describe a transition between a decelerated Universe to an accelerated one without the need for dark energy.
\end{abstract}

\section{Introduction}

General Relativity (GR) is one of the cornerstones of modern physics being stated as the standard
model of gravitation and cosmology. However, in the last years, shortcomings came out in the
Einstein's theory and the investigations whether GR is the fundamental theory capable of explaining
the gravitational interaction in different regimes initiated.

Combined data from Cosmic Microwave Background Radiation (CMB)~\cite{Wetterich1988} and from
Baryonic Acoustic Oscillations (BAO), indicate that the Universe is spatially flat, it is in
accelerated expansion~\cite{Riess1998, Riess2004}, and it is composed of $96\%$ of unknown
matter-energy, commonly known as dark matter and dark energy respectively. It is widely accepted
that the reason for the present accelerated expansion phase of the Universe is due dark
energy~\cite{Sahni2020, Bamba2012, Copeland2006}, while that an invisible matter (or dark matter)
accounts for the galaxies' rotation curves flatness~\cite{rubin/1980, rubin/1985}.

To overcome this situation different researchers came up with more sophisticated gravity theories by
modifying the Einstein-Hilbert action, which gave arise a new avenue known as modified or extended
theories of gravity. The extended theories of gravity have born out as an opportunity to solve
problems which are still without explanation within GR framework. The $f(R)$ theory of
gravity is one of the most well studied theories, and consists of choosing a more general
action to replace the Einstein-Hilbert one, assuming that the gravitational action is an arbitrary
function of the Ricci scalar $R$ as discussed in Refs.~\cite{Capozziello2002, Nojiri2003}.

In this work we are particularly interested in the $f(R,T)$ theory of gravity that is a
generalization of $f(R)$ type theories. The $f(R,T)$ gravity, proposed by Harko et
al.~\cite{Harko2011}, consists of choosing a gravitational action as an arbitrary function of the
Ricci scalar and also the trace of the energy-momentum tensor $T$. Moraes~\cite{Moraes2019} has used
$f(R,T)= R + f(T)$ to calculate the trace of the $f(R,T)$ gravity field equations. In this case, the
author describes only a minimal coupling between the Ricci tensor and an arbitrary function of the
energy-momentum tensor, i.e., a specific model. Here, we are going further in calculating the trace
of the $f(R,T)$ gravity field equations and also deriving a new field equation for the theory that does
not depend on the matter Lagrangian. In our approach matter and curvature can have a more complex
coupling, i.e., it is a general approach. As pointed by~\cite{Fisher2019Sep}, a more rich phenomenology could arise from a non-minimal geometry-matter coupling, what is within the motivations behind the present work.

This new general approach eliminates the Lagrangian ambiguity choice. We argue that the trace of the energy-momentum tensor is the macroscopic description of the more
fundamental gravity structure, i.e., it is the quantity that encodes the degree of freedom of the
matter to the scalar curvature. We show that our proposal resolves two limitations: the Lagrangian
choice, and the fact that the energy-momentum tensor cannot be the perfect fluid. From this novel approach we consider that the trace of the energy-momentum is an ``unknown'' variable, and thus, the trace of the field equations can be exploited to eliminate it.

This article is organized as follows: In Section~\ref{sec:frt} presents a basic overview on general
properties of the $f(R,T)$ gravity, in \S\ref{sec:traceless} we derive the traceless field
equations for a generic $f(R,T)$ functional, in \S\ref{sec:consistent} we present a consistent
approach for the Lagrangian ambiguity choice, in \S\ref{sec:application} we apply the theory to
  describe the matter era of cosmology and show that a transition between a decelerated Universe to
  an accelerated one is possible in $f(R,T)$ cosmology and finally in Section~\ref{sec:4} we conclude and discuss possible applications of the theory presented here.

\section{$f(R,T)$ gravity}\label{sec:frt}
The $f(R,T)$ gravity is derived by adopting the following gravitational action~\cite{Harko2011}
\begin{equation}\label{eq:1}
    S=\int {\rm d}^4 x \sqrt{-g} \left( \frac{f(R,T)}{16\pi}+\mathcal{L}\right),
\end{equation}
where $f(R,T)$ is a generic function of the Ricci scalar $R$, and to the trace $T$ of the
energy-momentum tensor $T_{\mu\nu}$. $\mathcal{L}$ represents the matter Lagrangian density. Natural
units are adopted and metric signature -2.

By variation of the action~\eqref{eq:1} with respect to the metric tensor $g_{\mu\nu}$, one obtains the field equations of the $f(R,T)$ gravity theory as follows
\begin{multline}\label{fe}
    f_R R_{\mu\nu}-\frac{1}{2} g_{\mu\nu}f + (g_{\mu\nu}\Box - \nabla_\mu \nabla_\nu )f_R = \\ 8\pi T_{\mu\nu} + f_T( T_{\mu\nu} - g_{\mu\nu}\mathcal{L}),
\end{multline}
where $\Box$ is the D'Alambertian
operator, $R_{\mu\nu}$ is the Ricci tensor and $\nabla_\mu$ represents the covariant derivative
associated with the Levi-Civita connection of $g_{\mu\nu}$. For sake of simplicity, we defined $f_R \equiv \frac{\partial{f(R,T)}}{\partial R}$ and $f_T \equiv \frac{\partial{f(R,T)}}{\partial T}$.

\section{Traceless $f(R,T)$ gravity}\label{sec:traceless}
Taking the trace of~\eqref{fe} we obtain
\begin{equation}\label{L}
   \mathcal{L}= \frac{f_T T- f_R R+2f-3 \Box f_R + 8\pi T}{4f_T}.
\end{equation}

Combining~\eqref{fe} with~\eqref{L} yields
\begin{multline}\label{traceless_fe}
    \left(R_{\mu\nu}- \frac{1}{4}g_{\mu\nu} R+\frac{1}{4}g_{\mu\nu}\Box - \nabla_\mu\nabla_\nu \right)f_R =\\ 8\pi T_{\mu\nu}- 2\pi g_{\mu\nu} T + f_T\left(T_{\mu\nu}-\frac{1}{4}g_{\mu\nu}T\right).
\end{multline}

The covariant derivative of the stress-energy tensor is given by
\begin{multline}\label{covaderiv}
    \nabla^\mu T_{\mu\nu}= \frac{f_T}{8\pi+f_T}\left[ (g_{\mu\nu}\mathcal{L}-T_{\mu\nu})\nabla^\mu {\ln}f_T \right. \\ \left.-\frac{1}{2}g_{\mu\nu}\nabla^\mu T +g_{\mu\nu}\nabla^\mu \mathcal{L}\right],
\end{multline}
where $\mathcal{L}$ can be eliminated from Eq.~\eqref{L}. As one can see, the four-divergence is
non-null and in a traceless formulation of the field equations, the $f(R,T)$ shares a similarity
with the unimodular gravity as will see ahead.

\subsection{$f(R,T)$ gravity and unimodular gravity, connection through energy-momentum violation}
Trying to deal with elementary particles in a geometrical framework, Einstein proposed~\cite{einstein/1919, einstein/1952} in
1919 a trace-free field equation

\begin{equation}\label{EFE-trace}
  R_{\mu \nu}-\frac{1}{4}g_{\mu \nu}R=8 \pi\left(T_{\mu \nu}-\frac{1}{4}g_{\mu \nu}T\right).
  \end{equation}

The formulation derived from the Einstein-Hilbert was persuaded in order to have an understanding in the right-hand side of the field
equations of General Relativity. The gravitational field equations involve only traceless parts of the Riemann/energy-momentum tensor.

  Nowadays, this formulation was reborn as ``unimodular gravity'', due to a fixation on the metric
  determinant -det$g_{\mu\nu}\equiv g=1$, and it is applied to solve the problem of the discrepancy
  between the vacuum energy density and the observed value of the cosmological
  constant~\cite{anderson/1971, ng/1990, ng/1991, ellis/2011, ellis/2013}.

  In Eq.~\eqref{EFE-trace}, the Bianchi identity still holds for the Einstein tensor, $\nabla^\mu G_{\mu\nu}=0$, but the vanishing of the four-divergence of energy-momentum tensor, $\nabla^\mu T_{\mu\nu}=0$, is not a geometrical consequence. As have been shown, the difference between the field equations in unimodular and in GR is a scalar stress $1/4(T+R/8\pi)g^{\mu\nu}$~\cite{anderson/1971}.

  The field equations are derived by restricting the variations preserving the volume form. These
  restrictions lead to violations of the energy-momentum conservation. For a conservative case, the
  condition

  \begin{equation}
    \nabla^{\mu}\left(8\pi T_{\mu\nu} - \frac{8\pi}{4}g_{\mu\nu}T - \frac{1}{4}g_{\mu\nu}R\right)=0,
  \end{equation}
  must be satisfied and it leads to GR with cosmological constant, i.e., dark energy.

  In the case
    of $f(R,T)$ gravity, which is a theory with a presence of coupling in the gravitational field, the non-vanishing of the energy-momentum tensor, Eq.~\eqref{covaderiv}, arises without
    restrictions in the variations and it is associated with particle creation in a
    quantum level, being plausible that gravitational field theories intrinsically contain effective particle
    creation in a phenomenological description~\cite{harko/2020}. Particle creation is
    a feature in quantum field theories described in curved spacetime and in noncommutative quantum
    field theories, which is field theory in a noncommutative spacetime and can be interpreted as a low
    energy limit of a quantum gravity theory. As we stated in a previous work~\cite{Lobato2019Apr}, the energy nonconservation in a four
    dimensional spacetime can be related to a noncommutative compact extra dimension with circular
    topology. In this regard, a letter by Josset, Perez \& Sudarsky~\cite{josset/2017} considered the unimodular gravity with violation of the
    conservation of energy-momentum, investigating sources of nonconservation in quantum
    mechanics. In a first scenario studied by them, is
    evoked a Markovian equation (used to describe creation and evolution of black holes) of the
    density matrix $\hat\rho$. This leads to a non-constant average energy $\langle{E}\rangle\equiv\Tr[\hat\rho, \hat H]$. In the second scenario the nonconservation arises naturally from quantum
    gravity. In a more recently letter~\cite{perez/2019}, exploring this second case, they showed
    that the nonconservation arises from the discreteness at the Planck level, similar to
    our line of thought~\cite{Lobato2019Apr}. They have shown that these quantum phenomena are
    relevant in a cosmological scale, i.e., the underline granularity of the spacetime would lead
    to the emergence of an effective dark energy. The relevance of the discreteness arises by the
    interaction of the gravity with scale-invariance-breaking fields (massive fields could interact
    with quantum gravity structure and exchange energy with it). The quantity that would describe
    macroscopically the phenomenon is the trace of the energy-momentum tensor $T$, which for a
    perfect fluid is given by $T = \rho - 3p$, the trace characterizes the breaking of the
    conformal and scale invariance~\cite{wess/1971}, and it is
    related to the scalar curvature, therefore captured geometrically by scalar curvature $R$. A non-vanishing of
    trace leads to a trace anomaly~\cite{schwimmer/2011, namavarian/2017}.

    As we can see, the
    trace is an important ingredient in the quantum and gravitational level description, and it is
    intrinsically associated with energy violations. We will use it in a more consistent approach to the Lagrangian ambiguity issue in $f(R,T)$ gravity in the next
    section.

\section{A more consistent approach to the Lagrangian ambiguity choice}\label{sec:consistent}

 In this section we present a new approach to the Lagrangian ambiguity problem in $f(R,T)$ gravity. Our
 approach consists of considering the trace of the energy-momentum tensor as a variable of the
 field equations.

 Taking the general definition~\cite{Fisher2019Sep} of the energy-momentum tensor given by

\begin{equation}
     T_{\mu\nu}=(p+\rho)u_\mu u_\nu +g_{\mu\nu}\mathcal{L},
 \end{equation}
 and calculating the trace we obtain that
 \begin{equation}\label{lagrangian2}
     \mathcal{L}=\frac{T-(p+\rho)}{4},
 \end{equation}
 then, the field equations become
\begin{multline}\label{newfeqs}
    f_R R_{\mu\nu}-\frac{1}{2} g_{\mu\nu}f + (g_{\mu\nu}\Box - \nabla_\mu \nabla_\nu )f_R  = \\ 8\pi T_{\mu\nu} + f_T\left( T_{\mu\nu} - \frac{g_{\mu\nu}}{4}(T-(p+\rho))\right).
\end{multline}
In this way, the field equations become independent of the matter Lagrangian. In a flat spacetime
limit, the trace is free of anomaly, however, considering a coupling, we can have trace anomaly,
i.e., correction terms to energy-momentum tensor, which would lead to phenomenological implications
as pointed by Perez \& Sudarsky~\cite{perez/2019}.

Rewriting the
energy-momentum tensor we have
\begin{equation}\label{tmunu}
    T_{\mu\nu}=(p+\rho)u_\mu u_\nu +g_{\mu\nu}\left(\frac{T-(p+\rho)}{4}\right).
\end{equation}

We can also calculate the four divergence of the energy-momentum tensor by
replacing~\eqref{lagrangian2} into Eq.~\eqref{covaderiv}. One must realize that, from now on, field
equations depend only on energy-momentum tensor and its trace, rather than matter Lagrangian. In previous works in $f(R,T)$ gravity
the trace of the energy-momentum
tensor depends on matter Lagrangian, being not well-defined. Assuming the trace to be an unknown entity, we can treat it as a variable of the $f(R,T)$ theory. To solve this issue one can take the trace of the field equations to obtain
\begin{equation}
    8\pi T +2f +f_T(p+\rho)= f_R R+ 3\Box f_R.
\end{equation}
When taking the trace of the field equations one more equation is added to the problems to be
solved. It is worth to quote that in this approach the trace, $T$, will have a similar role as the
curvature scalar, $R$, in $f(R)$ gravity theories. This approach has two major advances: it solves
the Lagrangian choice problem; and it also respects the fact that in $f(R,T)$ gravity the
energy-momentum tensor cannot be the one for perfect fluids. As the energy-momentum tensor is not well-defined as in GR our proposed approach solves this issue by coupling the trace of the field equation to themselves.

\section{Cosmology in the general approach for the Lagrangian ambiguity}\label{sec:application}
\subsection{Model I: $f(R,T)=R+\lambda T$}

Taking \eqref{newfeqs} and using $f(R,T)=R+\lambda T$, which is an extensively studied case, one obtains the field equations
\begin{equation}\label{narf}
    G_{\mu\nu}=(8\pi+\lambda)T_{\mu\nu} +g_{\mu\nu}\frac{\lambda}{4}(T+p+\rho).
\end{equation}

Therefore, the trace of the energy-momentum tensor becomes
\begin{equation}\label{Ttrace}
    T=-\frac{R+\lambda(p+\rho)}{8\pi+2\lambda},
\end{equation}
which states that the trace $T$ not only depends on pressure and energy density but it also depends on the curvature scalar, $R$.

Defining an effective energy-momentum tensor
\begin{equation}\label{eff}
    T_{\mu\nu {\ \rm I}}^{\rm eff}= \left(1+\frac{\lambda}{8\pi}\right)T_{\mu\nu}+\frac{\lambda}{32\pi}(T+p+\rho)g_{\mu\nu},
\end{equation}
we write the field equations in a compact form $G_{\mu\nu}=8\pi T_{\mu\nu\, I}^{\rm eff}$. Considering
the FLRW metric for these field equations we have
\begin{equation}\label{R}
R=-6(\dot{H}+2H^2)
\end{equation}
and
\begin{eqnarray}
&&3 H^{2}=8 \pi \rho^{\mathrm{eff}}_{\rm I}, \label{eq:camp1} \\
&&2 \dot{H}+3 H^{2}=-8 \pi p^{\mathrm{eff}}_{\rm I}, \label{eq:camp2}
\end{eqnarray}
where $H$ is the Hubble parameter.

Substituting \eqref{R} into \eqref{Ttrace}, and then this result into \eqref{eff}, we get

\begin{equation}
    \rho^{\rm eff}_{\rm I}=\frac{(3\lambda+24\pi)(p+\rho)+R}{32\pi},
  \end{equation}
  and
\begin{equation}
    p^{\rm eff}_{\rm I}=\frac{(\lambda+8\pi)(p+\rho)+R}{32\pi}.
\end{equation}

From \eqref{eq:camp1} and \eqref{eq:camp2}, we obtain
\begin{equation}\label{ev-prho}
    \dot{H}=-\frac{1}{2}(\lambda+8\pi)(p+\rho)
.\end{equation}
By assuming that pressure and energy density must be positive, equation \eqref{ev-prho} provides that, if $\lambda>-8\pi$ the Hubble parameter is decreasing, if $\lambda<-8\pi$, $H$ is increasing, and in case of $\lambda=-8\pi$, $H$ becomes a constant in time.

Considering the matter era ($p=0$) and isolating $\rho$ in both \eqref{eq:camp1} and \eqref{eq:camp2} and combining the results, one can arrive at
\begin{equation}\label{ev-time}
    \dot{H}=-2H^2,
\end{equation}
which solution is straightforward
\begin{equation}
    \frac{1}{H(t_0)}-\frac{1}{H(t)}=2(t_0-t).
\end{equation}
This equation gives a new model for the evolution of the Hubble constant with time, where $H(t)$ is noticeably a decreasing function.

By solving $a(t)=a_0\exp\left[\int_{t_0}^t Hdt \right]$ the solution for the scale parameter is
\begin{equation}
    a(t)=a_0(2H_0 t-2H_0 t_0 +1),
\end{equation}
which gives the deceleration parameter $q=-\ddot{a}a/\dot{a}^2=-8$. So, in this linear case, the theory does not predict a transition between a decelerated Universe
to an accelerated one. This indicates that a linear functional on $T$ (a well studied case in the
literature) cannot explain the observational data. However, the $f(R,T)$ theory is still possible
from the point of view of others functional, as we are going to show.

\subsection{Model II: $f(R,T)=R+\lambda T^2$}

Assuming now that $f(R,T)=R+\lambda T^2$, one can obtain the field equations as $G_{\mu\nu}=8\pi
T_{\mu\nu}^{\rm eff}$, where $T_{\mu\nu}^{\rm eff}$ reads now
\begin{equation}
    T_{\mu\nu {\ \rm II}}^{\rm eff}=\left(1+\frac{\lambda T}{4\pi}\right)T_{\mu\nu}+\frac{\lambda T}{16\pi}  (p+\rho)g_{\mu\nu}.
\end{equation}
In this case, the effective energy density and pressure are given by
\begin{equation}\label{rhoeff}
    \rho^{\rm eff}_{\rm II}=\left(1+\frac{\lambda T}{4\pi}\right)\left[(p+\rho)+\frac{T-(p+\rho)}{4}\right]+\frac{\lambda T}{16\pi} (p+\rho)
\end{equation}
\begin{equation}\label{peff}
    p^{\rm eff}_{\rm II}= -\left(1+\frac{\lambda T}{4\pi}\right)\left(\frac{T(p+\rho)}{4}\right)-\frac{\lambda T}{16\pi} (p+\rho).
\end{equation}

The modified Friedmann equations are
\begin{eqnarray}
&&3 H^{2}=8 \pi \rho^{\mathrm{eff}}_{\rm II}, \label{fr:camp1} \\
&&2 \dot{H}+3 H^{2}=-8 \pi p^{\mathrm{eff}}_{\rm II}. \label{fr:camp2}
\end{eqnarray}

To solve \eqref{fr:camp1} and \eqref{fr:camp2}, we use \eqref{peff} and \eqref{rhoeff} and assume
the matter era. Solving this numerically gives us the solutions for the Hubble parameter and energy
density, $\rho$. From the solution one can obtain the evolution of the scale parameter and, hence,
obtain the deceleration parameter, $q$. The initial conditions for the solutions used in this work
were: $H_0=67.4$ km s$^{-1}$ Mpc$^{-1}$ and $\rho_0=6\times 10^{-16}$ kg m$^{-3}$~\cite{aghanim/2020}.

\begin{figure}
    \centering
    \includegraphics[scale=0.5]{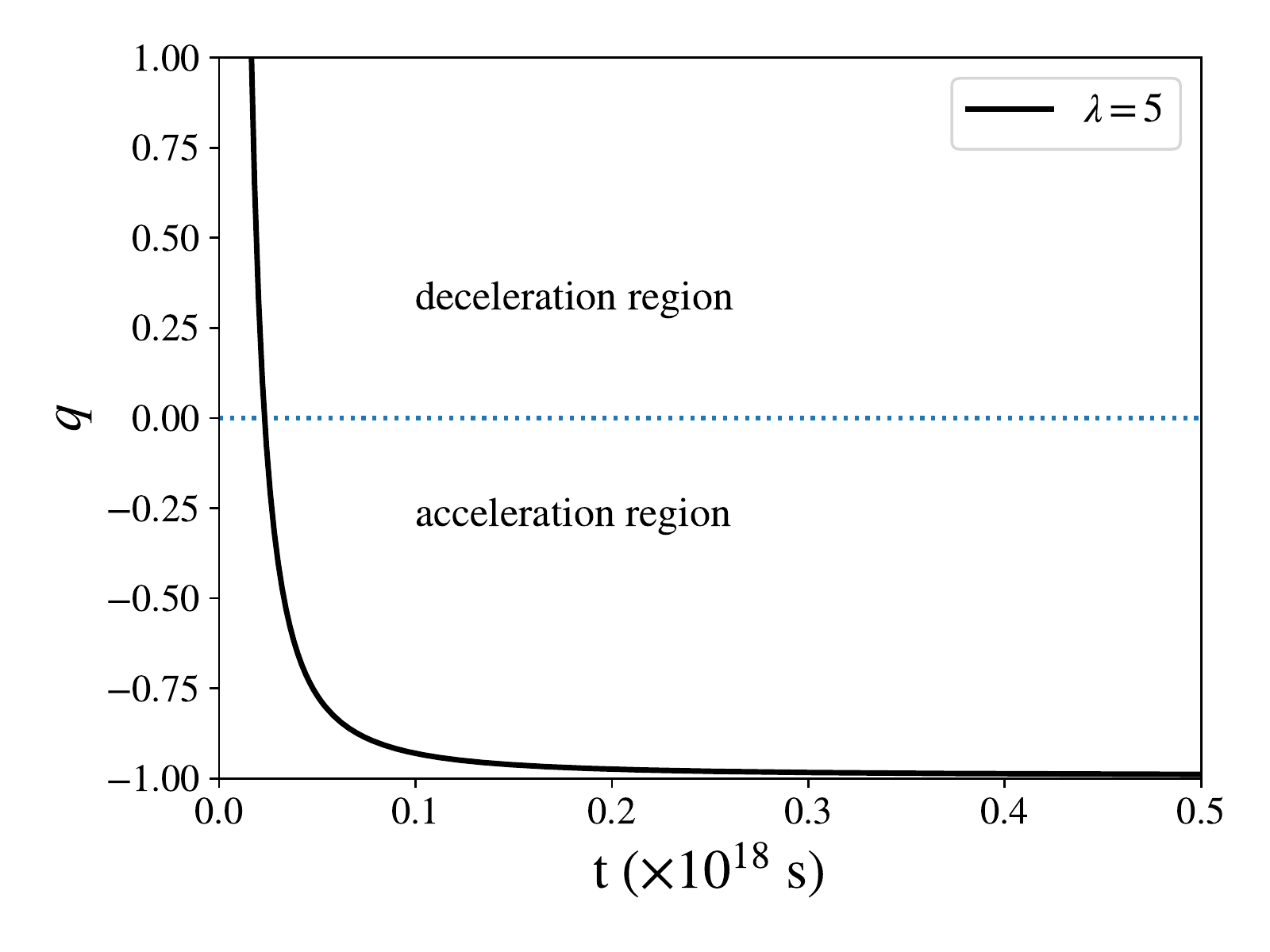}
    \caption{Evolution of the deceleration parameter with time. A change of signal on $q$ is
      presented, therefore, the theory, for $f(R, T)=R+\lambda T^2$, predicts a transition between a decelerated Universe to an accelerated one, without the need for dark energy.}
    \label{fig:1}
\end{figure}

The result for $q$ is shown on Fig. \ref{fig:1}, which indicates that the $f(R,T)$ gravity can
predict a transition between a decelerated Universe to an accelerated one for
$\lambda\approx1.6\pi$. In this second model, we have an agreement with the observational data,
which shows a transition between a decelerated Universe to an accelerated one.

\section{Discussion}\label{sec:4}

The $f(R,T)$ gravity has attracted a lot of researchers attention in the last few
years~\cite{Myrzakulov2012Nov, Alvarenga2013May, Houndjo2012Mar, Sharif2012Mar,
  Sharif2013Oct, Zubair2015Jun, Moraes2017Jul, Noureen2015Jul, Zaregonbadi2016Oct,
  Das2017Jun, Yousaf2017Jun, Sahoo2016Sep, Sharif2017Dec, Deb2018Apr, Yousaf2018Apr}. Nevertheless,
a few works have addressed the Lagrangian choice problem in the $f(R,T)$ gravity and modified
theories of gravity~\cite{Moraes2019, Fisher2019Sep, Faraoni2009Dec, Harko2010Feb, Avelino2018Mar}. In general, choices for matter Lagrangian among those works are $\mathcal{L}=\rho$ or $\mathcal{L}=-p$. In some works it is shown that Lagrangian may be an arbitrary function of pressure and energy density, $\mathcal{L}=\mathcal{L}(p,\rho)$, or considering an equation of state the dependence on pressure can be eliminated to give $\mathcal{L}=\mathcal{L}(\rho)$.

Moraes~\cite{Moraes2019} has provided a solution for the Lagrangian choice problem by deriving a field equation for the $f(R,T)$ gravity that does not depend on the matter Lagrangian. However, he considered the specific case $f(R,T)=R+f(T)$. The case studied by Moraes is an advance on $f(R,T)$ gravity research field, in the sense that now researchers have the possibility to study $f(R,T)$ gravity with no need for choosing a specific matter Lagrangian, thus, working on a general basis. In addition to Moraes' approach, we consider in this letter a generalization of his seminal idea. Here, we work with no specific case, so, the $f(R,T)$ functional remains as arbitrary as possible. This study was inspired by the work of Fisher \& Carlson~\cite{Fisher2019Sep}, where they studied the on-shell Lagrangian problem in $f(R,T)$ gravity. In their work, they suggest that only cross terms between matter and geometry could survive as a theory which brings new insights for the gravitational theory. Our work here is then presented as a possible way to eliminate the matter Lagrangian\footnote{for perfect fluids} as a variable of the field equations. This is done for any $f(R,T)$ functional, and hence it is also valid for cross terms between matter and geometry.

Another way to remove the matter Lagrangian form field equations is also presented here. Our
approach was again motivated by the work of Fisher and \& Carlson~\cite{Fisher2019Sep}, in the sense
that they have shown that the energy-momentum tensor cannot be given by the perfect fluid
definition. In this work, we take the general definition of the energy-momen\-tum tensor to remove
the dependence on matter Lagrangian of the field equations. We also argued that trace of
the energy-momentum tensor becomes an unknown variable that can be obtained from the trace of the field
equations. Hence, this approach unfolds two problems of the $f(R,T)$ gravity, which are the
Lagrangian choice one and the energy-momentum tensor that becomes not well-defined.

Finally, we have applied our results to cosmology, considering two specific cases: model I
and model II. In the case of model I, we have shown the cosmological inviability of the functional,
i.e., the linear case cannot explain a transition between a decelerated Universe to an accelerated
one. Nevertheless, for the case II, where we have $f(R,T)=R+\lambda T^2$, we found a viable
functional which predicts a transition between an decelerated era to an accelerated one, being in
agreement with the cosmological data without the introduction of dark energy.

Forthcoming applications of our approaches can be applied to flat rotation curves of galaxies,
astrophysical systems and so on. Others functional are also encouraged.

\begin{acknowledgements}
GAC thanks financial support from Coordena\c{c}\~ao de Aperfei\c{c}oamento Pessoal de N\'\i vel
Superior (CAPES), under projects: PDSE/88881.188302/2018--01 and PNPD/88887.368365/2019--00. FR and
HO would like to
thank CAPES for financial support. RVL has been supported by U.S. Department of Energy (DOE) under
grant DE-FG02-08ER41533, the LANL Collaborative Research Program by Texas A\&M System National Laboratory Office and Los Alamos National Laboratory.
\end{acknowledgements}

\bibliographystyle{spphys}
\bibliography{lib}

\begin{thebibliography}{10}
\providecommand{\url}[1]{{#1}}
\providecommand{\urlprefix}{URL }
\expandafter\ifx\csname urlstyle\endcsname\relax
  \providecommand{\doi}[1]{DOI \discretionary{}{}{}#1}\else
  \providecommand{\doi}{DOI \discretionary{}{}{}\begingroup
  \urlstyle{rm}\Url}\fi

\bibitem{Moraes2019}
P.H.R.S. Moraes, Eur. Phys. J. C \textbf{79}(8), 674 (2019)

\bibitem{Wetterich1988}
C.~Wetterich, Nucl. Phys. B \textbf{302}(4), 668 (1988)

\bibitem{Riess1998}
A.G. Riess, et~al., Astrophys. J. \textbf{116}(3), 1009 (1998)

\bibitem{Riess2004}
A.G. Riess, et~al., Astrophys. J. \textbf{607}(2), 665 (2004)

\bibitem{Sahni2020}
V.~Sahni, in \emph{The Physics of the Early Universe} (Springer, Berlin,
  Heidelberg, 2020), chap.~5, pp. 141--179

\bibitem{Bamba2012}
K.~Bamba, S.~Capozziello, S.~Nojiri, S.D. Odintsov, Astrophys. Space Sci.
  \textbf{342}(1), 155 (2012)

\bibitem{Copeland2006}
E.J. Copeland, M.~Sami, S.~Tsujikawa, Int. J. Mod. Phys. D \textbf{15}(11),
  1753 (2006)

\bibitem{rubin/1980}
V.C. Rubin, W.K. Ford, N.~Thonnard, The Astrophysical Journal \textbf{238}, 471
  (1980).
\newblock \doi{10/bhwfw5}

\bibitem{rubin/1985}
V.C. Rubin, D.~Burstein, W.K. Ford, N.~Thonnard, The Astrophysical Journal
  \textbf{289}, 81 (1985).
\newblock \doi{10/fkjbb5}

\bibitem{Capozziello2002}
S.~Capozziello, Int. J. Mod. Phys. D \textbf{11}(04), 483 (2002)

\bibitem{Nojiri2003}
S.~Nojiri, S.D. Odintsov, Phys. Rev. D \textbf{68}(12), 123512 (2003)

\bibitem{Harko2011}
T.~Harko, F.S.N. Lobo, S.~Nojiri, S.D. Odintsov, Phys. Rev. D \textbf{84}(2),
  024020 (2011)

\bibitem{Fisher2019Sep}
S.B. Fisher, E.D. Carlson, Phys. Rev. D \textbf{100}(6), 064059 (2019)

\bibitem{einstein/1919}
A.~Einstein, Sitzungsberichte der K{\"o}niglich Preu{\ss}ischen Akademie der
  Wissenschaften (Berlin), Seite 349-356.  (1919).
\newblock 00142

\bibitem{einstein/1952}
A.~Einstein, The Principle of Relativity. Dover Books on Physics. June 1, 1952.
  240 pages. 0486600815, p. 189-198 pp. 189--198 (1952).
\newblock 00095

\bibitem{anderson/1971}
J.L. Anderson, D.~Finkelstein, American Journal of Physics \textbf{39}(8), 901
  (1971).
\newblock \doi{10/cnwhq4}.
\newblock 00141

\bibitem{ng/1990}
Y.J. Ng, H.~{van Dam}, Physical Review Letters \textbf{65}(16), 1972 (1990).
\newblock \doi{10/btmw3v}.
\newblock 00048

\bibitem{ng/1991}
Y.J. Ng, H.~{van Dam}, Journal of Mathematical Physics \textbf{32}(5), 1337
  (1991).
\newblock \doi{10/b82d4h}.
\newblock 00072

\bibitem{ellis/2011}
G.F.R. Ellis, H.~van Elst, J.~Murugan, J.P. Uzan, Classical and Quantum Gravity
  \textbf{28}(22), 225007 (2011).
\newblock \doi{10/c57bqh}.
\newblock 00134

\bibitem{ellis/2013}
G.F.R. Ellis, General Relativity and Gravitation \textbf{46}(1), 1619 (2013).
\newblock \doi{10.1007/s10714-013-1619-5}.
\newblock 00074

\bibitem{harko/2020}
T.~Harko, F.S.N. Lobo, arXiv:2007.15345 [astro-ph, physics:gr-qc,
  physics:hep-th]  (2020)

\bibitem{Lobato2019Apr}
R.V. Lobato, G.A. Carvalho, A.G. Martins, P.H.R.S. Moraes, Eur. Phys. J. Plus
  \textbf{134}(4), 1 (2019)

\bibitem{josset/2017}
T.~Josset, A.~Perez, D.~Sudarsky, Physical Review Letters \textbf{118}(2),
  021102 (2017).
\newblock \doi{10/gfv7w7}.
\newblock 00083

\bibitem{perez/2019}
A.~Perez, D.~Sudarsky, Physical Review Letters \textbf{122}(22), 221302 (2019).
\newblock \doi{10/ggf4c3}

\bibitem{wess/1971}
J.~Wess, in \emph{Springer {{Tracts}} in {{Modern Physics}}, {{Volume}} 60},
  ed. by G.~H{\"o}hler, Springer {{Tracts}} in {{Modern Physics}} ({Springer},
  {Berlin, Heidelberg}, 1971), pp. 1--17.
\newblock \doi{10.1007/BFb0044910}

\bibitem{schwimmer/2011}
A.~Schwimmer, S.~Theisen, Nuclear Physics B \textbf{847}(3), 590 (2011).
\newblock \doi{10/cgvmjn}

\bibitem{namavarian/2017}
N.~Namavarian, Physical Review D \textbf{95}(10), 104015 (2017).
\newblock \doi{10.1103/PhysRevD.95.104015}

\bibitem{aghanim/2020}
N.~Aghanim, Y.~Akrami, M.~Ashdown, J.~Aumont, C.~Baccigalupi, M.~Ballardini,
  A.J. Banday, R.B. Barreiro, N.~Bartolo, S.~Basak, R.~Battye, K.~Benabed, J.P.
  Bernard, M.~Bersanelli, P.~Bielewicz, J.J. Bock, J.R. Bond, J.~Borrill, F.R.
  Bouchet, F.~Boulanger, M.~Bucher, C.~Burigana, R.C. Butler, E.~Calabrese,
  J.F. Cardoso, J.~Carron, A.~Challinor, H.C. Chiang, J.~Chluba, L.P.L.
  Colombo, C.~Combet, D.~Contreras, B.P. Crill, F.~Cuttaia, P.~de~Bernardis,
  G.~de~Zotti, J.~Delabrouille, J.M. Delouis, E.D. Valentino, J.M. Diego,
  O.~Dor{\'e}, M.~Douspis, A.~Ducout, X.~Dupac, S.~Dusini, G.~Efstathiou,
  F.~Elsner, T.A. En{\ss}lin, H.K. Eriksen, Y.~Fantaye, M.~Farhang,
  J.~Fergusson, R.~{Fernandez-Cobos}, F.~Finelli, F.~Forastieri, M.~Frailis,
  A.A. Fraisse, E.~Franceschi, A.~Frolov, S.~Galeotta, S.~Galli, K.~Ganga, R.T.
  {G{\'e}nova-Santos}, M.~Gerbino, T.~Ghosh, J.~{Gonz{\'a}lez-Nuevo}, K.M.
  G{\'o}rski, S.~Gratton, A.~Gruppuso, J.E. Gudmundsson, J.~Hamann, W.~Handley,
  F.K. Hansen, D.~Herranz, S.R. Hildebrandt, E.~Hivon, Z.~Huang, A.H. Jaffe,
  W.C. Jones, A.~Karakci, E.~Keih{\"a}nen, R.~Keskitalo, K.~Kiiveri, J.~Kim,
  T.S. Kisner, L.~Knox, N.~Krachmalnicoff, M.~Kunz, H.~{Kurki-Suonio},
  G.~Lagache, J.M. Lamarre, A.~Lasenby, M.~Lattanzi, C.R. Lawrence, M.L. Jeune,
  P.~Lemos, J.~Lesgourgues, F.~Levrier, A.~Lewis, M.~Liguori, P.B. Lilje,
  M.~Lilley, V.~Lindholm, M.~{L{\'o}pez-Caniego}, P.M. Lubin, Y.Z. Ma, J.F.
  {Mac{\'i}as-P{\'e}rez}, G.~Maggio, D.~Maino, N.~Mandolesi, A.~Mangilli,
  A.~{Marcos-Caballero}, M.~Maris, P.G. Martin, M.~Martinelli,
  E.~{Mart{\'i}nez-Gonz{\'a}lez}, S.~Matarrese, N.~Mauri, J.D. McEwen, P.R.
  Meinhold, A.~Melchiorri, A.~Mennella, M.~Migliaccio, M.~Millea, S.~Mitra,
  M.A. {Miville-Desch{\^e}nes}, D.~Molinari, L.~Montier, G.~Morgante, A.~Moss,
  P.~Natoli, H.U. {N{\o}rgaard-Nielsen}, L.~Pagano, D.~Paoletti, B.~Partridge,
  G.~Patanchon, H.V. Peiris, F.~Perrotta, V.~Pettorino, F.~Piacentini,
  L.~Polastri, G.~Polenta, J.L. Puget, J.P. Rachen, M.~Reinecke,
  M.~Remazeilles, A.~Renzi, G.~Rocha, C.~Rosset, G.~Roudier, J.A.
  {Rubi{\~n}o-Mart{\'i}n}, B.~{Ruiz-Granados}, L.~Salvati, M.~Sandri,
  M.~Savelainen, D.~Scott, E.P.S. Shellard, C.~Sirignano, G.~Sirri, L.D.
  Spencer, R.~Sunyaev, A.S. {Suur-Uski}, J.A. Tauber, D.~Tavagnacco, M.~Tenti,
  L.~Toffolatti, M.~Tomasi, T.~Trombetti, L.~Valenziano, J.~Valiviita, B.V.
  Tent, L.~Vibert, P.~Vielva, F.~Villa, N.~Vittorio, B.D. Wandelt, I.K. Wehus,
  M.~White, S.D.M. White, A.~Zacchei, A.~Zonca, Astronomy \& Astrophysics
  \textbf{641}, A6 (2020).
\newblock \doi{10/ggxrnb}

\bibitem{Myrzakulov2012Nov}
R.~Myrzakulov, Eur. Phys. J. C \textbf{72}(11), 1 (2012)

\bibitem{Alvarenga2013May}
F.G. Alvarenga, A.~de~la Cruz-Dombriz, M.J.S. Houndjo, M.E. Rodrigues,
  D.~S{\ifmmode\acute{a}\else\'{a}\fi}ez-G{\ifmmode\acute{o}\else\'{o}\fi}mez,
  Phys. Rev. D \textbf{87}(10), 103526 (2013)

\bibitem{Houndjo2012Mar}
M.J.S. Houndjo, O.F. Piattella, Int. J. Mod. Phys. D \textbf{21}(03), 1250024
  (2012)

\bibitem{Sharif2012Mar}
M.~Sharif, M.~Zubair, J. Cosmol. Astropart. Phys. \textbf{2012}(03), 028 (2012)

\bibitem{Sharif2013Oct}
M.~Sharif, S.~Rani, R.~Myrzakulov, Eur. Phys. J. Plus \textbf{128}(10), 1
  (2013)

\bibitem{Zubair2015Jun}
M.~Zubair, I.~Noureen, Eur. Phys. J. C \textbf{75}(6), 1 (2015)

\bibitem{Moraes2017Jul}
P.H.R.S. Moraes, R.A.C. Correa, R.V. Lobato, J. Cosmol. Astropart. Phys.
  \textbf{2017}(07), 029 (2017)

\bibitem{Noureen2015Jul}
I.~Noureen, M.~Zubair, A.A. Bhatti, G.~Abbas, Eur. Phys. J. C \textbf{75}(7), 1
  (2015)

\bibitem{Zaregonbadi2016Oct}
R.~Zaregonbadi, M.~Farhoudi, N.~Riazi, Phys. Rev. D \textbf{94}(8), 084052
  (2016)

\bibitem{Das2017Jun}
A.~Das, S.~Ghosh, B.K. Guha, S.~Das, F.~Rahaman, S.~Ray, Phys. Rev. D
  \textbf{95}(12), 124011 (2017)

\bibitem{Yousaf2017Jun}
Z.~Yousaf, M.~Ilyas, M.Z.u.H. Bhatti, Eur. Phys. J. Plus \textbf{132}(6), 1
  (2017)

\bibitem{Sahoo2016Sep}
P.K. Sahoo, B.~Mishra, P.~Sahoo, S.K.J. Pacif, Eur. Phys. J. Plus
  \textbf{131}(9), 1 (2016)

\bibitem{Sharif2017Dec}
M.~Sharif, A.~Siddiqa, Eur. Phys. J. Plus \textbf{132}(12), 1 (2017)

\bibitem{Deb2018Apr}
D.~Deb, B.K. Guha, F.~Rahaman, S.~Ray, Phys. Rev. D \textbf{97}(8), 084026
  (2018)

\bibitem{Yousaf2018Apr}
Z.~Yousaf, M.Z.u.H. Bhatti, M.~Ilyas, Eur. Phys. J. C \textbf{78}(4), 1 (2018)

\bibitem{Faraoni2009Dec}
V.~Faraoni, Phys. Rev. D \textbf{80}(12), 124040 (2009)

\bibitem{Harko2010Feb}
T.~Harko, Phys. Rev. D \textbf{81}(4), 044021 (2010)

\bibitem{Avelino2018Mar}
P.P. Avelino, L.~Sousa, Phys. Rev. D \textbf{97}(6), 064019 (2018)

\end{thebibliography}
\end{document}